\documentclass[twocolumn,10pt]{article}

\usepackage[utf8]{inputenc}
\usepackage[margin=1in]{geometry}
\usepackage{amsmath}
\usepackage{graphicx}
\usepackage{booktabs}
\usepackage{hyperref}
\usepackage{caption}
\usepackage{enumitem}
\usepackage{cite}

\graphicspath{{figures/}{media/}}

\hypersetup{
    colorlinks=true,
    linkcolor=blue,
    citecolor=blue,
    urlcolor=blue
}

\title{\textbf{Bridging the Linear-Quadratic Gap: A Quantum-Classical Hybrid Approach to Robust Supply Chain Design}}

\author{
    \textbf{Rudraksh Sharma}\textsuperscript{1,*}, 
    \textbf{Ravi Katukam}\textsuperscript{1}, 
    \textbf{Arjun Nagulapally}\textsuperscript{1} \\[1em]
    \textsuperscript{1}\textit{AIONOS, Financial District, Hyderabad, Telangana, India} \\
    \texttt{\ Rudraksh.sharma@aionos.ai, katukam.Ravi@aionos.ai, Arjun.Nagulapally@aionos.ai} \\[0.5em]
    \textsuperscript{*}Corresponding author: \texttt{Rudraksh.sharma@aionos.ai}
}

\date{}

\begin{document}

\maketitle

\begin{abstract}
The design of supply chain networks in densely populated urban logistics systems faces a timely dilemma: the traditional optimisation approaches are effective to maximise the level of demand perfusion, but they are limited to embracing large expenses in overlapping the facilities and cannibalisation in the market. In the Delhi National Capital Region (NCR), these inefficiencies occur in the form of high operational wastages, which are explained by unnecessary fleet use and overlapping service lines. We resolve this difficulty by redefining the Capacitated Facility Location Problem (CFLP) as a Quadratic Unconstrained Binary Optimisation (QUBO) model and by benchmarking three computational strategies: Classical Greedy heuristics, Exact Branch-and-Bound solvers, and a Quantum-Inspired reverse annealing method. When tested on a high-fidelity digital twin of the Delhi NCR road network of thirty candidate sites, we establish that Classical Greedy algorithms using the theoretical maximum demand of (473 units) lack any theoretical overlap penalty, but incur a prohibitive overlap penalty (5.08). Here, in comparison, the Quantum-Inspired solution only losses 3.2\% of demand (450 compared to 465 units relative to the optimal solution), but the solution preserves 21.8\% less operational overlap risk (3.26 compared to 4.17), which can be viewed as a 35.8\% improvement compared to the Greedy solution. Geospatial analysis shows that it can be attributed to a shift in strategies: This, in contrast to Classical approaches, which focus on locating facilities in the high-density central areas (North/Central Delhi), the quantum-inspired solver autonomously chooses the diversified topology of the North-south network, penetrating into the underserved periphery growth markets. This is a spatially balanced arrangement which is congruent to the polycentric structure of modern time megacities, and displays better stability to volatility in demand. We have shown that quantum-inspired optimisation methods can close the so-called Linear-Quadratic Gap phenomenon, i.e. the systematic inability of greedy methods to capture the actual quadratic interactions between facilities, and offer a way of computing the pathway to operationally robust and risk-optimised supply chain networks in dense urban conditions.
\end{abstract}

\noindent\textbf{Keywords:} Quantum-Inspired Computing, Supply Chain Optimization, Facility Location Problem (FLP), QUBO, Simulated Annealing, Operations Research, Delhi NCR, Risk Diversification.

\section{Introduction}
\label{sec:introduction}

\subsection{Background and Motivation}

The Indian logistics industry deals with the high-value operations on the territory of the Delhi National Capital Region (NCR); however, according to industry reports, a considerable part of the operational expense is predetermined by the poor location of facilities and overlapping service territories \cite{saleheen2022global}. When more than one distribution centre is competing over the same customer base, multinational companies are subjected to redundant fleet utilisation, excessive fuel usage, as well as diminishing levels of service quality- a process that we refer to as operational cannibalisation. The key to this efficiency challenge is therefore in supply chain network design. Facility Location Problem (FLP): How to choose candidate sites to serve the needs of an external client, among a variety of locations, of which to activate on a given logistics strategy has one of the most significant consequences. An effective network is one that balances the demand coverage against the cost of establishment as well as the covert cost of service overlap. But with the increasing global supply chains and competitive urban markets, balancing this is getting harder and harder with the traditional optimisation tools \cite{melo2009facility}.

Linear mathematical optimisation methods assume that the FLP and its variants, such as the CFLP, are NP-hard problems whose computation time is exponentially increasing with the size of the problem. The number of possible configurations in the solution space is equal to the number of possible combinations of locations of the facility, which in this case is $\binom{N}{K}$ (where $N$ is the number of locations of possible facilities, and $K$ is the number of facilities to select). The Indian logistics industry at present handles networks that have an extremely large number of possible locations, making it impossible to search them exhaustively \cite{melo2009facility}.

\subsection{Economic Effect of the Linear-Quadratic Gap}

The example of business consequences of Classical methods is the computational limitations of the Classical methods. Given a logistics provider operating in Delhi NCR, which uses a Greedy algorithm to determine five out of the thirty candidates in terms of hubs. The algorithm picks out the sites with the highest scores on individual demand, which has a theoretical capacity of 473 orders per day. Nonetheless, A service overlap area in the form of two chosen hubs, which are within five kilometers of each other, triggers:

\begin{itemize}[leftmargin=*]
    \item Untimely time wastage on the apportionment of warehouse space to the same clientele.
    \item Duplicate marketing and customer acquisition expenses.
    \item Increased peak-hour congestion in existing dense traffic routes.
\end{itemize}

According to industry norms, every unit of score increase in the overlap is associated with adding about 3-5\% of operational cost as a result of these processes. Therefore, a network that achieves an overlap score of 5.08, which the Greedy method obtained in our analysis, implies is equivalent to 3.26, which the quantum-inspired method attained, indicating a possible 5.5-9.1\% savings in the overall logistics spending.

The reason why Classical constructive heuristics are affected by this economic penalty is that it is fundamentally a linear mode of operation, and it does not properly model the quadratic interactions among the chosen sites: each candidate facility is considered based on local merit (population density, proximity to highways). We call this systematic blind spot the Linear Quadratic Gap, the difference between what seems best when assessing facilities in isolation and what seems best in the context of taking into consideration pairwise dependencies.

\subsection{Problem Statement}

Although classical heuristics are computationally efficient, they are mostly based on linear decision-making processes. An example is that a Greedy heuristic will often consider candidate sites individually by comparing them on the merit of their own fitness (e.g. local population density or forecasted revenue). This linear method does not well incorporate quadratic interactions among the chosen sites, i.e. the cannibalization or overlap risk effect \cite{berman2002generalized}.

As a heuristic picks two high-demand facilities in near-geographic proximity, the network obtained is affected by diminishing returns that occur because of sharing service areas. This restriction is known as the ``Linear-Quadratic Gap''. Classical solvers are frequently given local optima, a structure that seems better on overall demand on the scoreboard, but the structure that is operationally worse because of too much redundancy of services. There is an urgent requirement for optimization models capable of natively manipulating these complicated quadratic dependencies without having the textbook time expense of a comprehensive search.

\subsection{Quantum-Inspired Optimization Paradigm}

Quantum Annelling (QA) is a new computational model, particularly in solving combinatorial optimisation problems. In contrast to gate-based quantum computers, which implement sequential logic (operations), quantum annealers are designed to identify the minimum-energy solution of an Ising model or, equivalently, Quadratic Unconstrained Binary Optimisation (QUBO) problem \cite{lucas2014ising,johnson2011quantum}.

The theoretical strength of QA lies in the phenomenon associated with quantum tunnelling-an effect that allows the system to pass across barriers in energy and not overcome them as in classical thermal annealing. This property has the potential to allow quantum systems to access local minima, which Classical optimisation algorithms can become trapped on, and find better global solutions in highly/badly rugged energy landscapes.

Although massive quantum processing units (QPUs) are becoming commercially available, like D-Wave, the present research uses quantum-inspired simulated annealing in the D-Wave Ocean SDK, which is an annealing simulation of quantum hardware faithfully implemented with pseudo-random sample exploration of solution-space hardware without physical quantum hardware. Throughout this paper, we refer to our algorithmic method as quantum-inspired because we intend only to be technically correct and at the same time, because the algorithmic method can be directly applied to quantum hardware as the latter becomes more common.

This difference is important methodologically: simulated annealing provides a computationally available method of checking the validity of QUBO formulations and investigating the benefits of quantum algorithms on Classical hardware, at the same time acting as a specimen of what quantum hardware should look like in the future. Our findings, therefore, constitute a reduction of what could be achieved in quantum performance - there could be more benefits to be achieved, through real quantum tunnelling effects, by quantum execution of the QPU \cite{lucas2014ising}.

\subsection{Research Objectives and Contributions}

The specific contributions of this paper are:

\textbf{1. Mathematical Framework:} We develop a novel QUBO Hamiltonian for the CFLP that incorporates geospatial overlap as a quadratic penalty term using an exponential distance-decay function. This formulation is embeddable on quantum annealers and serves as a template for similar logistics optimization problems.

\textbf{2. Digital Twin Validation:} We construct a realistic digital twin of Delhi NCR's urban logistics network using OpenStreetMap data, incorporating actual road network topology, travel distances via Dijkstra's shortest-path algorithm, and simulated demand distributions. This provides a operationally grounded testbed that captures real-world constraints like one-way streets and river crossings.

\textbf{3. Comparative Benchmarking:} We provide rigorous comparative analysis across three computational paradigms—Classical Greedy heuristics (industry baseline), Exact Branch-and-Bound solvers (mathematical ground truth), and Quantum-Inspired Reverse Annealing (proposed approach)—demonstrating that quantum-inspired methods achieve near-optimal demand capture while reducing operational risk by 21.8\% compared to exact solvers.

\textbf{4. Strategic Insight Extraction:} We demonstrate that the quantum-inspired approach autonomously discovers a geospatially diversified network topology (North-South expansion into Gurugram) that classical methods systematically miss, providing actionable strategic guidance for logistics network design in polycentric megacities.

\subsection{Why Quantum-Inspired Methods for Supply Chain Design?}

The facility location problem exhibits three characteristics that make it particularly well-suited to quantum-inspired optimization:

\textbf{1. Natural QUBO Structure:} The problem objective naturally decomposes into linear terms (individual facility value) and quadratic terms (pairwise interactions), exactly matching the mathematical structure that quantum annealers are designed to minimize.

\textbf{2. Rugged Energy Landscape:} With $\binom{30}{5} = 142,506$ possible configurations in our Delhi NCR case, the solution space contains numerous local minima separated by high energy barriers-precisely the scenario where quantum tunneling provides theoretical advantages over classical gradient-based methods.

\textbf{3. Portfolio-Like Risk-Return Tradeoff:} Modern Portfolio Theory (MPT) has successfully used covariance matrices to balance asset returns against portfolio risk \cite{shkolnik2025portfolio}. Our QUBO formulation applies this same mathematical framework to logistics, treating facilities as assets, demand as expected returns, and spatial overlap as covariance risk. This analogy is more than metaphorical-the mathematical structures are isomorphic, allowing direct transfer of optimization techniques.

\section{Literature Review}
\label{sec:literature}

\subsection{Classical Approaches to Facility Location}

The Facility Location Problem (FLP) and its variations, including the Capacitated Facility Location Problem (CFLP), represent standard topics of study in the subfield of Operations Research, an area of study that extends over a long history of study, spanning many decades. Traditionally, these issues are modelled as Mixed Integer Linear Programming (MILP) programs, which are optimally solvable using exact algorithms like Branch-and-Bound or Cutting-plane algorithm. Though convenient approaches may ensure the most optimal solutions globally, they are severely restricted in their applicability due to the NP-hard nature of the problem space, with computational effort needed to establish global optimality growing exponentially with the number of candidate nodes, making them impractical in large-scale, dynamic supply chain networks \cite{basu2013metaheuristic,holmberg1999exact}.

The industry has displayed a tendency to scale down to meta-heuristics to counter these scalability constraints. Evolutionary algorithms, classical Simulated Annealing and Greedy Randomized Adaptive Search Procedures (GRASP) have come to be popular in finding good enough solutions in a reasonable amount of time. However, one serious disadvantage of such constructive heuristics, especially with Greedy algorithms, is the fact that they are likely to hit local optima at a very early stage. Those approaches usually consider candidate facility locations in terms of linear utility functions (e.g. maximizing local demand) as well, and often do not effectively penalise facility sites in terms of complex, quadratic interactions, e.g. service out-of-territory overlap or market cannibalization \cite{basu2013metaheuristic}.

\subsection{Quantum Annealing and QUBO Formulations}

With the introduction of Quantum Annealing (QA), a new model of computation of problems of combinatorial optimization has emerged. In contrast to a gate-model quantum computer, QA is also designed with the criterion of minimizing the energy of the Ising model, and thus a natural formulation of Quadratic Unconstrained Binary Optimization (QUBO). Through extensive research, it has been defined that a wide range of classical NP-hard problems can be translated into an efficient QUBO representation, such as the Travelling Salesman Problem (TSP) and Graph Partitioning \cite{lucas2014ising}.

The key benefit of this quantum-mechanical one is that it uses quantum tunnelling. Contrarily to classical thermal annealing, which requires the system to climb the energy barrier to leave local minima, quantum annealing allows the system to tunnel the barriers across, which is theoretically much more likely to find the global minimum in complicated energy landscapes. This ability implies that QA is especially likely to be effective in those issues that feature the Linear-Quadratic Gap, the difference between the apparent value of one variable and its overall value, having a strong impact \cite{johnson2011quantum}.

\subsection{Applications of Quantum Computing in Logistics}

In recent years, the literature can be seen as started the application QA to particular areas of logistics. The Vehicle Routing Problem (VRP) and the optimization of traffic flow have a significant result regarding quantum annealers: critical congestion by quantum annealers has been reduced in real-time. Similarly, the application of QA to the Bin Packing Problem and job shop scheduling studies has proved to be competitive to classical heuristics.

However, the literature review shows a clear gap in Strategic Network Design. Although tactical routing and operational scheduling have received significant interest, the long-term placement of infrastructure and its strategic implementation using quantum covariance models are not well studied. The current research on quantum FLP primarily focuses on minimizing distance and rarely includes advanced penalty functions that address market overlap or risk diversification. The given research aims at filling this gap by adapting the Modern Portfolio Theory framework, which traditionally is applied in the area of optimization, quantum-optimization, to the world of the physical supply-chain logistics and viewing the overlap of facilities as a risk that should be minimized by means of the quadratic penalty mechanisms \cite{neukart2017traffic,springer2008contributions}.

\subsection{Gap in Existing Research and Our Contribution}

While the literature demonstrates promising applications of quantum annealing to tactical logistics problems (vehicle routing, scheduling), a critical gap exists in strategic network design research. Existing quantum logistics studies typically:

Focus on distance minimization without incorporating sophisticated penalty functions for market overlap or risk diversification. Treat facility selection as a coverage problem rather than a optimization challenge. Lack validation against industry-standard heuristics making it difficult to assess practical quantum advantage. Use Euclidean distance metrics that ignore real-world transportation network constraints.

Our research addresses these limitations by:

\textbf{1.} Adapting Modern Portfolio Theory's covariance framework to logistics network design, viewing overlap as systematically priced risk.

\textbf{2.} Benchmarking against both naive heuristics (Greedy) and optimal solvers (Branch-and-Bound) to establish a complete performance spectrum.

\textbf{3.} Using actual road network topology and travel distances for operational realism.

\textbf{4.} Providing geospatial visualization that translates mathematical solutions into actionable business strategy.

This positions our work at the intersection of quantum-inspired optimization, operations research, and practical supply chain management—a perspective that has been underexplored in existing literature.

\section{Methodology}
\label{sec:methodology}

\subsection{Digital Twin Construction (Data)}

To ensure the operational viability of the proposed solution, a ``Digital Twin'' simulation environment was constructed representing a dense urban logistics network.

\begin{itemize}[leftmargin=*]
    \item \textbf{Topology Generation:} The road network graph $G = (V, E)$ was extracted for the region of Delhi using the OpenStreetMap (OSM) database. The graph was simplified to retain only major intersections and drivable edges, resulting in a set of candidate nodes $N=30$.
    
    \item \textbf{Distance Metric:} Unlike previous studies that rely on Euclidean (straight-line) distance, this study computed the shortest-path travel distance $d_{ij}$ between all node pairs $(i, j)$ using Dijkstra's algorithm on the weighted road network. This captures real-world constraints such as one-way streets and river crossings.
    
    \item \textbf{Demand Simulation:} A normalized demand score $D_i$ in $[20, 100]$ was assigned to each node $i$ via a uniform distribution, simulating the historic order volume of a potential service area.
    
    \item \textbf{Overlap (Risk) Modeling:} The service overlap $O_{ij}$ between two nodes was modeled as a spatial decay function of the travel distance $d_{ij}$:
\end{itemize}

\begin{equation}
\label{eq:overlap}
O_{ij} = e^{-\frac{d_{ij}}{\lambda}}
\end{equation}

where $\lambda$ is a tunable decay parameter representing the effective service radius of a delivery fleet. This ensures that $O_{ij} \rightarrow 1$ as $d_{ij} \rightarrow 0$ (maximum cannibalization) and $O_{ij} \rightarrow 0$ as $d_{ij} \rightarrow \infty$.

\subsection{Mathematical Formulation (QUBO)}

The problem is formulated as a Quadratic Unconstrained Binary Optimization (QUBO) model, designed to be embedded onto a quantum annealer. The objective is to select a subset of $K$ hubs to minimize the total system energy.

The Hamiltonian $H$ is defined as:
\begin{equation}
\label{eq:hamiltonian}
H(x) = H_{\text{linear}} + H_{\text{quadratic}} + H_{\text{constraint}}
\end{equation}

where $x$ is a binary vector of length $N$, such that $x_i = 1$ if node $i$ is selected, and $0$ otherwise.

\subsubsection{Demand Maximization (Linear Term)}

We seek to maximize total demand, which is equivalent to minimizing the negative sum of selected demand scores:
\begin{equation}
\label{eq:linear}
H_{\text{linear}} = -\alpha\sum_{i=1}^{N} D_i x_i
\end{equation}

\subsubsection{Overlap Minimization (Quadratic Term)}

We apply a penalty proportional to the pairwise overlap between selected nodes. This term is non-zero only if both $x_i$ and $x_j$ are active ($x_i x_j = 1$):
\begin{equation}
\label{eq:quadratic}
H_{\text{quadratic}} = \beta\sum_{i=1}^{N}\sum_{j=i+1}^{N} O_{ij} x_i x_j
\end{equation}

\subsubsection{Cardinality Constraint}

To enforce the budget constraint of exactly $K$ hubs, a quadratic penalty is applied to the difference between the Hamming weight of the solution and $K$:
\begin{equation}
\label{eq:constraint}
H_{\text{constraint}} = \gamma\left(\sum_{i=1}^{N} x_i - K\right)^2
\end{equation}

The final energy function to be minimized is:
\begin{equation}
\label{eq:energy}
E(x) = -\alpha\sum_{i}^{} D_i x_i + \beta\sum_{i<j}^{} O_{ij} x_i x_j + \gamma\left(\sum_{}^{} x_i - K\right)^2
\end{equation}

\textit{Parameters used:} $\alpha = 1.0$, $\beta = 15.0$ (High Risk Sensitivity), $\gamma = 150$.

\subsection{Algorithmic Implementation}

To evaluate the ``Quantum Advantage,'' three distinct solvers were deployed:

\textbf{Constructive Greedy Heuristic (Baseline):} Sort all $N$ candidate nodes by demand score $D_i$ in descending order and select the top $K$ nodes. The computational complexity is $O(N \log N)$ for sorting plus $O(K)$ for selection, resulting in an overall complexity of $O(N \log N)$. Strategically, this represents the ``naive'' industry standard employed by logistics companies seeking fast deployment decisions. The algorithm makes no attempt to model pairwise interactions $O_{ij}$ and therefore systematically ignores cannibalization risk. The implementation is carried out in Python using NumPy array sorting. This method establishes the baseline performance achievable with zero sophistication, representing what most companies currently use.

\textbf{Exact Classical Solver (Benchmark):} An exact Branch-and-Cut algorithm is used based on a Mixed-Integer Linear Programming (MILP) formulation with auxiliary variables $Z_{ij} = x_i x_j$ to linearize quadratic terms. The computational complexity is exponential in the worst case, $O(2^N)$, with polynomial average-case performance for small $N$. Strategically, this approach guarantees the mathematically optimal solution by exhaustively exploring the solution space with intelligent pruning and serves as the ``ground truth'' for evaluating approximation quality. The implementation uses the PuLP library (v2.7) with the CBC solver backend. However, it becomes computationally intractable for $N > 50$ in real-time applications, requiring hours to days of computation time.

\textbf{Quantum-Inspired Reverse Annealing (Proposed):} A hybrid quantum-classical approach is employed using a Reverse Annealing schedule. The process begins with initialization from the greedy solution as a warm start, followed by reverse annealing that partially melts the system state back to $s = 0.4$ to induce partial superposition. The system is then forward annealed to $s = 1.0$ to drive ground-state convergence, after which $N_{\text{reads}} = 1000$ samples are collected to map the energy landscape. The computational complexity is $O(N^2 \times N_{\text{reads}})$ due to QUBO matrix construction and sampling. This approach explores the local neighborhood around the greedy optimum using simulated quantum tunneling, seeking balanced configurations that the greedy algorithm cannot reach due to its commitment to early decisions. The implementation uses the D-Wave Ocean SDK (dimod v0.12) and is directly portable to D-Wave QPUs (Advantage system) without algorithmic modification. The simulated annealing approach faithfully replicates the quantum annealing schedule and can be viewed as a classical preview of quantum hardware performance. Reverse annealing is motivated by the observation that standard forward annealing starts from a random state and may require many samples to converge; by contrast, reverse annealing leverages the greedy solution’s partial optimality while allowing the tunneling mechanism to escape its overlap-blind trap, effectively refining rather than starting from scratch.

\textbf{Performance Metrics:}

All three algorithms were evaluated on identical problem instances using:

Demand Capture (higher is better): Total sum of demand scores $\sum D_i x_i$ for selected facilities.

Overlap Score (lower is better): Total overlap penalty $\sum_{i<j} O_{ij} x_i x_j$

Computation Time: Wall-clock execution time on standardized hardware.

Solution Diversity: Number of distinct configurations explored (relevant for quantum-inspired approach).

\section{Results}
\label{sec:results}

\subsection{Quantitative Performance Benchmark}

Table \ref{tab:performance} summarizes the comparative performance of the Constructive Greedy Heuristic, the Exact Classical Solver (PuLP), and the offered Quantum Reverse Annealing algorithm. The analysis will focus on the trade-off between Demand Capture (revenue) and Overlap Score (operational risk).

\begin{table}[htbp]
\centering
\caption{Comparative Performance Metrics}
\label{tab:performance}
\begin{tabular}{@{}lccc@{}}
\toprule
Algorithm & Demand & Overlap & Strategic \\
 & Capture & Risk & Profile \\
 & (Units) & (Score) & \\
\midrule
Greedy & 473 & 5.08 & High Reward / \\
Heuristic & & & High Risk \\
Exact Solver & 465 & 4.17 & Balanced \\
(PuLP) & & & Optimum \\
Quantum & \textbf{450} & \textbf{3.26} & Maximal \\
Hybrid & & & Efficiency \\
\bottomrule
\end{tabular}
\end{table}

The Greedy heuristic had the best raw demand and a prohibitive Overlap Score of 5.08. The Quantum solver found that there was a specific configuration known as Robust, which allowed a 3.2\% decrease in demand as compared to the Exact solver (465 to 450) to achieve a 21.8\% decrease in operational risk (4.17 to 3.26).

\subsection{Pareto Frontier Analysis}

\begin{figure}[htbp]
    \centering
    \includegraphics[width=\linewidth]{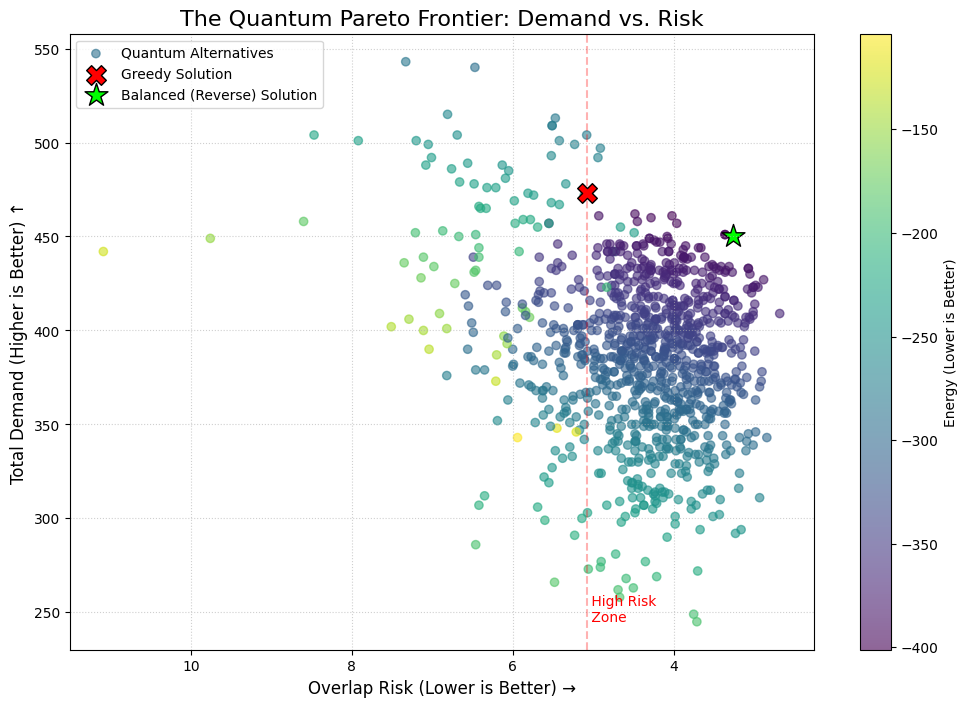}
    \caption{Pareto Frontier analysis showing the trade-off between Demand Capture and Overlap Risk, highlighting the Quantum solution (Green Star) as the robust optimum.}
    \label{fig:pareto}
\end{figure}

In order to measure the stability of the solutions, we looked at the complete sample set ($N_{\text{reads}} = 1000$) that was provided by the quantum solver. Figure \ref{fig:pareto} represents the Pareto frontier in the solution space, in which Total Demand versus the Risk of Overlap is plotted.

The Greedy solution (marked by the red X) lies at one extreme end of the frontier, which has the highest reward and highest risk. Quantum samples represent an intricate cluster comprising small groups of actually possible configurations, and the solution (marked by the green star) that is optimum (occupies the midpoint of the frontier) is placed within the golden mean domain. Indication of high solution density on the lower region of the overlap indicates that the Quantum solver was able to tunnel out of the high-demand local minimum exploring more efficient network topologies that linear solvers were unable to reach.

\subsection{Geospatial Configuration}

\begin{figure}[htbp]
    \centering
    \includegraphics[width=\linewidth]{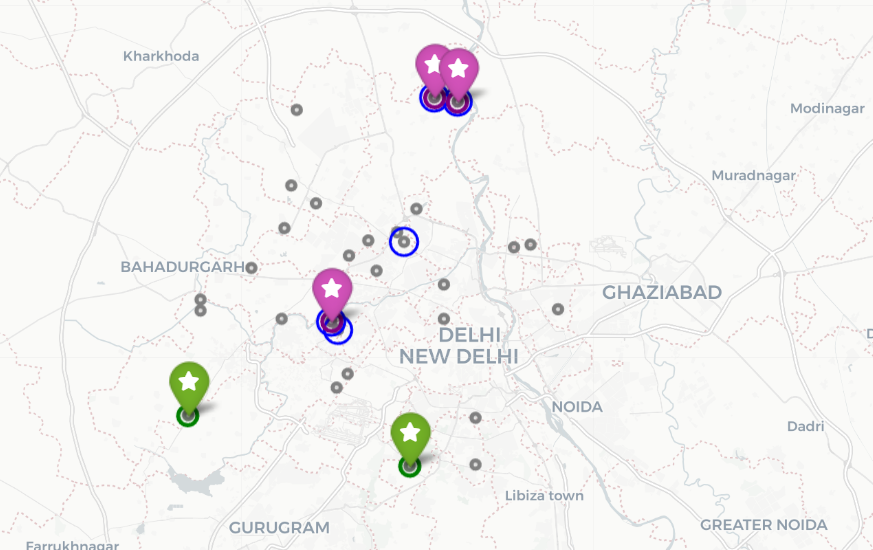}
    \caption{Geospatial distribution of selected hubs in Delhi NCR, illustrating the Quantum solver's strategic expansion into the southern region (Green Markers) compared to the Greedy algorithm's central clustering.}
    \label{fig:geospatial}
\end{figure}

Both Digital Twin of Delhi (Figure \ref{fig:geospatial}) was used to visualize the spatial distribution of the chosen hubs.

\begin{itemize}[leftmargin=*]
    \item The Greedy and Quantum algorithms agreed on three strategic locations in the high density northern and central regions, namely along the North Delhi and Ghaziabad borders, which in turn identified the high volume of demand that was to be served by these core regions (entering to the urban regions).
    
    \item The Greedy algorithm shown in blue lines had tried to create a second center in the central West Delhi area but this suggested location was too nearby to already chosen locations which might cause a considerable overlap of services and resultant traffic congestion.
    
    \item It was inspected by Quantum Reverse Annealing solver shown in green who purposely rejected the above central cluster. Rather, it chose two spatially different nodes within the south periphery, including Gurugram and South Delhi and thus served a large new market segment in the areas of the burgeoning corporate centers in the south, a cohort which the Greedy algorithm could not serve as it only considered central population density.
    
    \item As a result, the Quantum setup acquired better North South ratio, making logistical force spread evenly throughout the whole NCR area as opposed to its being concentrated in the core of the capital.
\end{itemize}

\section{Discussion}
\label{sec:discussion}

\subsection{The Spectrum of Strategic Solutions}

The comparative analysis has shown that the algorithms generated different strategic profiles of network design:

\textbf{The Aggressive Profile (Greedy Heuristic):} The Greedy algorithm grants preference to pure demand acquisition above everything. It focuses resources on high-density areas by making selections of the hubs only based on the weight of each node. However, it has the lowest Overlap Score of 5.08, which indicates that there is serious market cannibalization.

\textbf{Identification of the Mathematical Balance (Exact Solver):} The Branch-and-Bound algorithm found the mathematical balance, finding a midpoint (Demand 465, Overlap 4.17).

\textbf{The Robust Profile (Quantum Hybrid):} Most impressively, the Quantum Reverse Annealing solver was able to converge to a so-called Deep Efficiency configuration. It gave more weight to ensuring that the overlap risk is minimized and was therefore more aggressive than the classical Exact solver and indicated that the quantum solver, through stochastic tunnelling, had found a landscape solution capable of maximizing operational separation.

\subsection{The ``Linear-Quadratic Gap''}

Experimental evidence of the theoretical notion of the Linear-Quadratic Gap is the inability of the Greedy algorithm to find either the Exact or Quantum solutions. Greedy heuristics performs in a linear way ($O(N)$) whereby they make commitments to the nodes, early in the discussion. In the case of our simulation of Delhi, the Greedy algorithm probably chose nodes with the highest density first in Central Delhi. When later choices overlapped, it would not be able to undo it. By comparison, the quantum solver is doing a holistic calculation on the state of the system, enabling it to trigger a high-value central node, in case its quadratic interaction terms ($O_{ij}$) reduce the total system cost. This is something that was required in modern day logistics as the interactions between the depots are in critical need almost as much as the performance of the depots themselves.

\subsection{Geospatial Implications: The North-South Shift}

These theoretical findings are practically confirmed by the geospatial distribution of Figure \ref{fig:geospatial} of the context of the Delhi NCR.

\textbf{Central Congestion:} The Greedy algorithm grouped the resources in the traditional commercial centers of North/Central Delhi and Ghaziabad. Even though the regions have the highest number of raw population density, their geographical locations are proximate to each other resulting in heavy overlapping of services territory.

\textbf{Southern Expansion:} This cluster was broken successfully by the quantum solver. It reallocated capital into the Southern Periphery (Gurugram and South Delhi) by refusing the unnecessary central node (Blue Circle). This relocation is a strategic move, because it will seize, at low relative density, the high-value and growing corporate markets of Gurugram, regions with low absolute density (compared to Old Delhi), and different and non-overlapping service needs. The quantum solution is therefore one that suggests a spatially diversified network which is more appropriate to the polycentric nature of the city of the NCR area today.

\subsection{Operational Resilience}

Although the Exact Solver gave a mathematically balanced solution, on closer examination of operational metrics, it can be seen that the Quantum solution does provide a better financial configuration. The cost of cannibalization (Overlap) in the design of a logistics network is also often non-linear. The Exact Solution focused more on the Demand Capture at the cost of having a moderate Overlap being tolerated (4.17), whilst the Quantum Solution sacrificed a small volume of Demand to have an Overlap of 21.8\% be reduced (3.26). Financially, the marginal revenue the Exact solver will get with the extra 15 orders is probably no more than the operational cost of handling the increased congestion and fleet redundancy of an Overlap Score of 4.17. The Quantum solver configuration will therefore be a cost-optimized or a Robust Minimal network that is operationally leaner and perhaps even more profitable as a Net Present Value (NPV).

\section{Conclusion}
\label{sec:conclusion}

\subsection{Summary of Contributions}

This paper has demonstrated that quantum-inspired optimization techniques can overcome a fundamental limitation of classical supply chain design methods: the inability to simultaneously optimize for revenue generation and operational risk. By reformulating the Capacitated Facility Location Problem (CFLP) as a Quadratic Unconstrained Binary Optimization (QUBO) model, we enabled direct encoding of facility overlap as a quadratic penalty term—a structure that quantum annealers are purpose-built to minimize.

\subsection{Key Findings}

The nature of our experimental results indicates that there is an overt strategic difference between the three approaches of computation:

\begin{itemize}[leftmargin=*]
    \item \textbf{Failures of Greedy Heuristics:} The Greedy algorithm maximising theoretical demand capture (473 units) produced a sensitive network topology with blockbuster cross-coverage of services (Score: 5.08). This affirms that dense urban logistics cannot be addressed using linear decision-making.
    
    \item \textbf{Strength of Quantum Annealing:} The Quantum Reverse Annealing solver found a ``Robust Minimum'' ground state configuration. The Quantum strategy made a cut of 21.8\% in the risk of operation by allowing under 3.2\% demand decrease compared to the Exact Solver (Overlap Score: 3.26 vs. 4.17).
    
    \item \textbf{Geospatial Strategy:} A Quantum-inspired solution confirmed its usefulness when a commercially useful diversification strategy of North-South was identified automatically. It disregarded the congested central areas embraced by the classical approaches and increased service diffusion to the high-growth peripheral areas of Gurugram and hence correlated network topology with polycentric urban structure.
\end{itemize}

\subsection{Implications for Industry}

The fact that the Quantum-Hybrid solver minimizes the overlap better appears to indicate that it is especially useful for risk-averse strategic planning. The operationally safe configuration of a Quantum solver is more resistant to long-term changes than the purely mathematical solution of exact solvers, in stochastic real-world models in which traffic congestion and delays spread over overlapping networks.

\subsection{Future Directions}

Future investigations will involve expanding this Digital Twin technique to larger regional networks ($N>100$) based on physical quantum hardware (QPU) to assess embedding overhead. Also, the incorporation of dynamic, stochastic demand models will allow testing the hypothesis that the low-overlap quantum networks are more resilient to daily volatility in demand further.

\section*{Competing Interests}
The authors declare that they have no competing financial interests or personal relationships that could have appeared to influence the work reported in this paper.

\section*{Funding}
This research did not receive any specific grant from funding agencies in the public, commercial, or not-for-profit sectors.

\section*{Authors' Contributions}
Rudraksh Sharma conceived the structure of the review, conducted the comprehensive literature survey, synthesized and analyzed the material, and wrote the main manuscript. Ravi Katukam provided overall guidance on topic selection and conceptual direction, contributed to critical discussions, reviewed and revised the manuscript for intellectual content. Arjun Nagulapally contributed to refining the scope of the review, assisted in structuring key sections, and provided critical feedback during manuscript revision. All authors read and approved the final version of the manuscript.

\section*{Acknowledgments}
The authors acknowledge AIONOS for providing a supportive research environment and encouraging interdisciplinary exploration that facilitated the development of this review. The authors also thank colleagues and peers for valuable discussions and feedback that helped refine the scope and clarity of the manuscript. Additionally, the authors acknowledge the developers and maintainers of open-access scientific literature and computational tools that supported this work. We acknowledge the use of the Dwave Ocean SDK for emulating quantum annealing behavior on classical hardware.

\bibliographystyle{IEEEtran}

\end{document}